# Experimental study of extrinsic spin Hall effect in CuPt alloy

Rajagopalan Ramaswamy[1], Yi Wang[1], Mehrdad Elyasi[1], M. Motapothula[2], T. Venkatesan[1,2,3,4,5], Xuepeng Qiu[6] and Hyunsoo Yang[1,2,*]

[1]Department of Electrical and Computer Engineering, National University of Singapore, 117576, Singapore

[2]NUSNNI-Nanocore, National University of Singapore, 117411, Singapore

[3]Department of Physics, National University of Singapore, Singapore 117542, Singapore

[4]Department of Materials Science and Engineering, National University of Singapore, Singapore 117542, Singapore

[5]Integrated Science and Engineering Department, National University of Singapore, Singapore 117542, Singapore

[6]Institute of Solid State Physics and School of Physics Science, Tongji University, Shanghai 200092, China

*eleyang@nus.edu.sg

We have experimentally studied the effects on the spin Hall angle due to systematic addition of Pt into the light metal Cu. We perform spin torque ferromagnetic resonance measurements on Py/CuPt bilayer and find that as the Pt concentration increases, the spin Hall angle of CuPt alloy increases. Moreover, only 28% Pt in CuPt alloy can give rise to a spin Hall angle close to that of Pt. We further extract the spin Hall resistivity of CuPt alloy for different Pt concentrations and find that the contribution of skew scattering is larger for lower Pt concentrations, while the side-jump contribution is larger for higher Pt concentrations. From technological perspective, since the CuPt alloy can sustain high processing temperatures and Cu is the most common metallization element in the Si platform, it would be easier to integrate the CuPt alloy based spintronic devices into existing Si fabrication technology.

# I. INTRODUCTION

The spin Hall effect (SHE) exploits spin-orbit (SO) interaction in the non-magnetic material (NM) to convert an unpolarized charge current into a pure spin current and vice-versa [1-3]. The mechanism of SHE eliminates the need of a ferromagnetic spin polarizer or an external magnetic field to electrically generate spin currents. The spin currents thus generated by SHE have been utilized to efficiently manipulate the magnetization of an adjacent ferromagnet using SO induced torques [4-6]. The spin current generation efficiency of a material by SHE is quantified by the spin Hall angle ($\theta_{SH}$) and it is desired to have a large $\theta_{SH}$ for constructing efficient spintronics devices.

Since the origins of SHE lie in SO coupling and the strength of SO coupling is expected to be larger for heavier elements, SHE has been widely explored in heavy metals such as Pt, Ta, W and Hf [4-13]. SHE in such heavy metals can be understood to arise from intrinsic SHE mechanism which results from the effects of SO interaction on the electronic band structure [14-17]. However, in complementary metal oxide semiconductor (CMOS) technology, the commonly utilized elements for metallization are Cu and Al, which have a very small $\theta_{SH}$. Studies have been carried out by adding nonmagnetic impurities with strong SO interactions, such as Bi, Ir, and Pb, in Cu [18-22] to enhance the magnitude of $\theta_{SH}$ through the extrinsic SHE mechanisms. Such extrinsic SHE mechanisms rely on electron scattering on the impurity centers and are of two kinds, namely skew scattering [23] and side-jump [24]. The advantage of extrinsic SHE mechanisms is that the $\theta_{SH}$ can be tuned by changing the relative concentrations of host and impurity atoms [25-27].

In this work, we study the effects on $\theta_{SH}$ due to the addition of Pt into the light metal Cu. The CuPt alloy has been predicted to have a sizable extrinsic spin Hall effect [28-33]. The $\theta_{SH}$ of the CuPt alloy, with different Pt concentrations, is estimated using the technique of spin-torque ferromagnetic resonance (ST-FMR) [5]. We find that as the concentration of Pt in the CuPt alloy increases, the $\theta_{SH}$ of the system linearly increases. From the analysis of different contributions to the spin Hall resistivity in the CuPt alloy, we find that for lower Pt concentrations ($< \sim 12.7\%$) the contribution of skew scattering is larger than that of side-jump, while for higher Pt concentrations ($> \sim 12.7\%$) the contribution of side-jump is larger than that of skew-scattering.

## II. EXPERIMENTAL DETAILS

The film stack structure for the ST-FMR measurements is Si substrate/Py (5)/Cu$_{1-x}$Pt$_x$ (6)/MgO (1)/SiO$_2$ (3) (nominal thickness in nm), where Py is Permalloy (Ni$_{81}$Fe$_{19}$) and x (0 − 100%) is the atomic ratio of Pt in Cu$_{1-x}$Pt$_x$ alloy, determined using Rutherford backscattering spectroscopy. The entire film stack was deposited onto a thermally oxidized Si substrate at room temperature using magnetron sputtering with a base pressure of $< 2 \times 10^{-9}$ Torr. The composite alloy of Cu$_{1-x}$Pt$_x$ was deposited by co-sputtering Cu and Pt targets. In order to tune the Pt concentration (x) in the Cu$_{1-x}$Pt$_x$ alloy, the sputtering power of Cu was fixed at 120 W and the sputtering power of Pt was varied from 0 to 150 W for x < 75%; while the sputtering power of Pt was fixed at 60 W and the sputtering power of Cu was varied between 0 and 60 W for x ≥ 75%.

The deposited films were subsequently patterned into rectangular microstrips of dimensions $L$ $(130 \ \mu m) \times W$ $(20 \ \mu m)$ using optical photolithography and Ar ion milling. In the subsequent step, a coplanar waveguide (CPW) was fabricated using optical photolithography and sputter deposition to make electrical contacts with the microstrip devices. The gap (G) between ground and signal electrodes of the CPW was varied in the range $35 - 90\,\mu m$ among the different devices on a film in order to tune the device impedance close to $\sim 50\,\Omega$. Figure 1(a) shows the 3D schematic of our devices with a measurement configuration for ST-FMR. Figure 1(b) shows a microscope image of the ST-FMR device. An orange circle is drawn in Fig. 1(b) to indicate the region of the device illustrated in Fig. 1(a). All the ST-FMR measurements in this study were carried out at room temperature.

For ST-FMR measurements, a microwave current of a fixed frequency ($f$ = 7, 8, or 9 GHz) is applied to Py/Cu$_{1-x}$Pt$_x$ bilayer. Simultaneously, an external magnetic field $H_{ext}$ is applied at an angle $\theta_H$ = 35° with respect to the current channel (see Fig. 1(a)). Due to the principle of SHE, the oscillating charge current in the non-magnetic CuPt alloy is converted into a transverse oscillating spin current which in turn exerts an oscillating spin Hall torque on the ferromagnet (Py). The a.c. charge current in the CuPt layer also generates an alternating Oersted field induced torque on the Py layer. These oscillating torques induce magnetization precession in the Py layer and the resistance of the bilayer oscillates due to the anisotropic magnetoresistance effect at the same frequency as the magnetization precession. Consequently, a d.c. mixing voltage $V_{mix}$ is produced by the product of applied alternating charge current and oscillating resistance. A nanovoltmeter is used to measure the $V_{mix}$ signal across the device. For a given frequency $f$ of the microwave current, $H_{ext}$ is swept to meet the resonance condition given by the Kittel's relation. The ST-FMR spectrum is obtained by measuring $V_{mix}$ as a function of $H_{ext}$. Figure 1(c) shows representative ST-FMR spectra from Py/Cu$_{1-x}$Pt$_x$ bilayers for x = 0, 6.6, 13.7, and 19.7% at a microwave frequency of 8 GHz and a nominal microwave power of 16 dBm.

## III. RESULTS AND DISCUSSION

The ST-FMR spectra can be fitted using the equation, $V_{mix} = V_S F_S \left( H_{ext} \right) + V_A F_A \left( H_{ext} \right)$, [5] where $F_S \left( H_{ext} \right)$ is a symmetric Lorentzian function of amplitude $V_S$ and $F_A \left( H_{ext} \right)$ is an antisymmetric Lorentzian function of amplitude $V_A$. Figure 2 shows the fitting (black curve) of the ST-FMR spectra at 8 GHz for the negative $H_{ext}$ range for x = 0, 6.6, 13.7, and 19.7%. The red and green curves in Fig. 2 correspond to the symmetric and antisymmetric components of the fitting, respectively, for the different x. We could observe that as the concentration of Pt increases, the amplitude of symmetric Lorentzian function increases. The Oersted field induced torque from the charge current in CuPt layer is out-of-phase with the magnetization precession and thus generates an antisymmetric Lorentzian spectrum about the resonance field, while the spin Hall torque from the generated spin current is in-phase with the magnetization precession and hence produces a symmetric Lorentzian spectrum about the resonance field. Therefore, the increase of the symmetric component of ST-FMR line shape indicates that the spin Hall torque on Py increases due to the generation of a larger spin current density in CuPt, as the Pt concentration in CuPt alloy increases.

$\theta_{SH}$ is the ratio of spin current density generated in the NM for a given charge current density. Therefore, $\theta_{SH}$ can be expressed to be proportional to the ratio $V_S / V_A$ according to the equation $\theta_{SH} = (V_S / V_A)(e\mu_0 M_S t d / \hbar)[1 + (4\pi M_{eff} / H_{ext})]^{1/2}$, where $M_S$ and $M_{eff}$ are the saturation and effective magnetization of Py layer, respectively [5]. $t$ and $d$ are the thicknesses of Py and CuPt alloy, respectively. Figure 3(a) shows the extracted values of $\theta_{SH}$ of $Cu_{1-x}Pt_x$ alloy from the $V_S/V_A$ ratio (blue circles) averaged from ST-FMR data for three frequencies (7, 8, and 9 GHz). From Fig. 3(a), we could observe that as the Pt concentration increases, the $\theta_{SH}$ of CuPt monotonically increases until x = ~33%. Furthermore, we could observe that a CuPt alloy with 28% Pt can give rise to a $\theta_{SH}$ of 0.054, which is comparable to $\theta_{SH}$ of 0.055 obtained from pure Pt [5,34-38].

The $V_S/V_A$ ratio method utilized to determine $\theta_{SH}$ values assumes that the field-like torque arises from Oersted field only and not from SO effects. However, if the SO effects generate a significant field-like torque term [39,40] and thus contribute to $V_A$ [41-43], the value of $\theta_{SH}$ may not be accurately estimated using the $V_S/V_A$ ratio method. In order to eliminate such an issue, $\theta_{SH}$ can be determined by considering only the symmetric component $V_S$ of the ST-FMR spectrum using the equation $\theta_{SH} = \dfrac{\sigma_{SHE}}{\sigma} = \dfrac{1}{\sigma}\left(\dfrac{4V_S M_s t \Delta}{E I_{rf} \gamma cos\theta_H \left(dR/d\theta_H\right)}\right)$, [35,41,42] where $\Delta$ is the linewidth of the Lorentzian ST-FMR spectrum, $E$ and $I_{rf}$ are the microwave electric field and current through the device, respectively, $dR/d\theta_H$ is angular dependent magnetoresistance of the device at $\theta_H = 35°$, and $\sigma_{SHE}$ and $\sigma$ are the spin Hall and longitudinal charge conductivities of the CuPt alloy, respectively. Figure 3(a) also shows the values of $\theta_{SH}$ for different x extracted from only $V_S$ (red squares). We observe that extracted values of $\theta_{SH}$ from $V_S/V_A$ ratio and only $V_S$ are almost similar indicating that the field-like torque from the SO effects is negligible compared to the Oersted field induced torque. Furthermore, we have also evaluated the spin pumping induced inverse SHE voltage in our devices and found it to be negligible (see supplementary material [44]).

Apart from $\theta_{SH}$, we have also extracted the effective Gilbert damping coefficient, $\alpha_{eff}$, from the ST-FMR measurements, using the relation, $\Delta = H_0 + 2\pi\alpha_{eff} f / \gamma$. Figure 3(b) shows extracted $\alpha_{eff}$ for different Pt concentrations. We observe that in the Cu rich regime the extracted $\alpha_{eff}$ remains relatively constant, which may arise from the saturating behavior of the effective spin mixing conductance ($g_{\uparrow\downarrow}^{\text{eff}}$, which is proportional to $\alpha_{eff}$) as observed before in the case of CuIr [45]. As we increase the Pt content, the interface between CuPt and Py changes from Cu-rich to Pt-rich regime. It is known that, compared to the Py/Cu bilayer, the Py/Pt bilayer offers a larger Gilbert damping enhancement due to a larger spin mixing conductance [5,46-49] and/or enhanced magnetic proximity effect [50]. Therefore, we observe an increase in the $\alpha_{eff}$ as the Pt concentration is increased beyond ~50%. Further, from Fig. 3(b), the extracted $\alpha_{eff}$ in the Cu-rich regime is ~0.01 which is ~2 times smaller than $\alpha_{eff}$ extracted from ST-FMR measurements in Py/Pt bilayer [5,36,37]. Thus, in addition to its significant $\theta_{SH}$, CuPt alloy offers a smaller damping enhancement in Py which makes it suitable for applications requiring a lower Gilbert damping [18,51].

Figure 3(c) shows the plot of the longitudinal resistivity for CuPt alloy for a thickness of 6 nm ($\rho_{CuPt}$) as a function of Pt concentration with a fit using the parabolic relation governed by the Nordhiem rule for homogenous solid solutions [52,53], $\rho_{CuPt} = C_1 X (100 - X) + \left(\frac{\rho_{Pt}}{100}\right) X + \left(\frac{\rho_{Cu}}{100}\right)(100 - X)$, where $C_1$ is proportionality constant for the parabolic term, $\rho_{Pt}$ and $\rho_{Cu}$ are values of the longitudinal resistivity for pure Pt and pure Cu, respectively, of thicknesses 6 nm and $X$ is equivalent to x expressed in percentage. We observe that, except for a deviation at x ≈ 60%, the data fits well to the above equation. A sudden drop in $\rho_{CuPt}$ at x ≈ 60% may arise due to changes in preferential distribution of Pt atoms near the equiatomic concentration as observed before [54]. Nevertheless, the majority of the data follow the parabolic Nordhiem relation indicating that our CuPt alloy is homogeneous for most of the Pt concentrations, at least in the Cu-rich regime (x < 50%). We restrict our further analyses in the Cu-rich regime.

In order to identify the different contributions of SHE in CuPt, we first plot $\theta_{SH}$ as a function of $\rho_{CuPt}$, in the Cu-rich regime, as shown in Fig. 4(a). It is observed that $\theta_{SH}$ increases linearly with respect to $\rho_{CuPt}$. However, both intrinsic and side-jump give rise to a linear contribution in $\theta_{SH}$ with respect to $\rho_{CuPt}$, and thus it is not straightforward to identify the different contributions of SHE directly from Fig. 4(a). Therefore, we first isolate the intrinsic and extrinsic contributions to the spin Hall resistivity using the following equation [55]

$$-\rho_{SH} = \sigma_{SH}^{int}\rho_{CuPt}^2 - \rho_{SH}^{imp} \tag{1}$$

where $\rho_{SH}$ is the total spin Hall resistivity of CuPt alloy determined from relation $\theta_{SH} = (-\rho_{SH} / \rho_{CuPt})$, $\sigma_{SH}^{int}$ is the intrinsic contributions of Cu to the spin Hall resistivity, and $\rho_{SH}^{imp}$ is the extrinsic spin Hall resistivity induced by the Pt. In Eq. (1), the contributions of phonons to the total spin Hall resistivity are not considered as they are negligible [55-58]. However, we have not neglected the contributions of $\sigma_{SH}^{int}$ in our analyses, due to a non-zero $\theta_{SH}$ in Cu even though it is one order of magnitude smaller than that in CuPt alloy. To determine $\sigma_{SH}^{int}$, we consider the case x = 0%, for which $\rho_{SH}^{imp} = 0$ and $\rho_{CuPt} = \rho_{Cu}$. Hence, $\sigma_{SH}^{int}$ can be written as $\sigma_{SH}^{\text{int}} = -\rho_{SH} / \rho_{Cu}^2 = \theta_{SH,Cu} / \rho_{Cu}$ , where $\theta_{SH,Cu}$ is the $\theta_{SH}$ of Cu (x = 0%). Substituting the expressions for $\sigma_{SH}^{\text{int}}$ and $\rho_{SH}$ into Eq. (1), we obtain the following equation

$$-\rho_{SH}^{imp} = \rho_{CuPt}\theta_{SH} - (\theta_{SH,Cu} / \rho_{Cu})\rho_{CuPt}^2 . \qquad (2)$$

Figure 4(b) shows the plot of $\left| \rho_{SH}^{imp} \right|$ (calculated using Eq. (2)) as a function of $\rho_{imp}$, where $\rho_{imp}$ is given by the relation, $\rho_{imp} = \rho_{CuPt} - \rho_{Cu}$ . We then fit the data in Fig. 4(b) to the relation $\left| \rho_{SH}^{imp} \right| = \theta_{SH}^{SS}\rho_{imp} + \sigma_{SH}^{SJ}\rho_{imp}^2$ , [19,25,55,59,60] where $\theta_{SH}^{SS}$ and $\sigma_{SH}^{SJ}$ are the contributions of skew scattering and side-jump to the extrinsic SHE induced by Pt. From the fitting, we extract the following values: $\theta_{SH}^{SS} = 0.022 \pm 0.006$ and $\sigma_{SH}^{SJ} = 0.0014 \pm 0.0001~\mu\Omega^{-1}~cm^{-1}$ . For these extracted values of the $\theta_{SH}^{SS}$ and $\sigma_{SH}^{SJ}$ , we find that for $\rho_{imp} = \theta_{SH}^{SS} / \sigma_{SH}^{SJ} = 15.7~\mu\Omega~cm$ , the contributions from skew scattering and side-jump to extrinsic SHE are equal. Therefore, for $\rho_{imp} < 15.7~\mu\Omega~cm$, the skew scattering contribution to extrinsic SHE is larger than the side-jump contribution, while for $\rho_{imp} > 15.7~\mu\Omega~cm$ the side-jump contribution is larger. From Fig. 3(c), the value of $\rho_{imp} = 15.7 \mu\Omega~cm$ corresponds to x = ~12.7%. Hence, in the CuPt system, for low Pt concentrations (< 12.7%) the skew scattering contribution to extrinsic SHE is larger, while for higher Pt concentrations (> 12.7%) the side-jump contribution is larger, which agrees well with the previous reports [27,29,59,60]. Further, in the case of the CuPt, we observe that the sign of $\theta_{SH}^{SS}$ and $\sigma_{SH}^{SJ}$ are the same and positive. On the other hand, in the case of n-GaAs [61], it was observed that the contributions of skew scattering and side-jump are opposite. However, it is to be noted that the sign of skew scattering is sensitive to the nature of impurity atoms [30,61] and thus, depends on particular material combination. Further, the same sign of skew scattering and side-jump obtained from CuPt is similar to the results obtained in Gd based alloys in the context of anomalous Hall effect [60].

Apart from the Pt concentration dependence, we have also measured $\theta_{SH}$ as a function of the thickness ($d$) of CuPt layer for some compositions (x = 3.5, 13.7, and 19.7%), as shown in Fig. 5(a). For each composition, we fit the data using $\theta_{SH}(d) = \theta_{SH}^0 (1 - \mathrm{sech}(d / \lambda_{SH}))$, [3,5,62] to extract the intrinsic spin Hall angle ($\theta_{SH}^0$) and spin diffusion length ($\lambda_{SH}$). The fittings are shown as dashed lines for each composition and the $\lambda_{SH}$ is indicated. It can be observed that our room temperature value of $\lambda_{SH}$ (~ 2 nm) is 4–5 times smaller than experimentally reported values of 8 ± 2 nm and 11 ± 3 nm in $Cu_{94}Pt_6$ alloy at 4.2 K [32,33]. This difference is expected since it is known that the room temperature value of $\lambda_{SH}$ can be 3–4 times smaller than that measured at low temperatures (< 10 K) [55,63,64]. Figure 5(b) shows a plot of the $\lambda_{SH}$ as a function of the CuPt conductivity ($\sigma$) for a thickness of 6 nm and it is observed that the data can be fitted with a straight line. The linear variation of $\lambda_{SH}$ with $\sigma$ suggests that the spin relaxation in the CuPt alloy in the considered concentrations could arise from the Elliot-Yafet mechanism [65]. The product $\rho_{CuPt}\lambda_{SH}$ obtained in the considered concentrations is in the range of $0.63-0.69\,f\Omega\,m^2$, which is comparable to the reported values for the case of pure Pt ($0.58-0.77\,f\Omega\,m^2$) [38,49,65].

Table I summarizes the extracted room temperature values of $\theta_{SH}^0$, $\lambda_{SH}$ and the product $\theta_{SH}^0\lambda_{SH}$ (which is the figure of merit for inverse SHE) for the CuPt alloy for x = 3.5, 13.7, and 19.7% and compares them with the corresponding reported values (at 10 K) for CuBi, CuPb and CuIr alloys. It can be observed the CuPt alloy has smaller values of $\theta_{SH}^0\lambda_{SH}$ compared with the other alloys, which may suggest that the other Cu based alloys are better choice for inverse SHE detection. However, the CuPt alloy exists as a single phase solid solution for temperatures upto ~1000°C due to high solubility of Pt in Cu [66,67]. On the other hand, the solubility in the case of CuBi, CuPb, and CuIr alloys is restricted to less than ~1% Bi, ~ 0.5% Pb and ~10% Ir, respectively [67-69]. Furthermore, the CuBi and CuPb alloys cannot be annealed beyond ~300 °C, due to a low melting point of Bi and Pb and the solubility of Bi and Pb in Cu also degrades upon annealing [68,69]. Therefore, compared to the other Cu alloys, it would be easier to integrate CuPt alloy based spintronic devices into the existing CMOS platform as CuPt alloy can sustain CMOS backend processing temperatures, such as 400°C (refer supplementary material [44] for estimated $\theta_{SH}$ from annealed CuPt alloy).

## IV. CONCLUSIONS

We have investigated the $\theta_{SH}$ in CuPt alloy, and find that $\theta_{SH}$ increases as the Pt concentration increases in the Cu-rich regime. By analyzing the different contributions to extrinsic SHE mechanism, we find that the contribution of skew scattering is larger than side-jump for lower Pt concentrations (< 12.7%), while the contribution side-jump mechanism is larger for higher Pt concentrations. We also find that the $Cu_{72}Pt_{28}$ alloy is as efficient as Pt in spin current generation but with a smaller damping enhancement. Compared with other Cu based alloys, it would be easier to integrate CuPt alloy into the existing CMOS platform since Cu is the most widely used metallization element and the CuPt alloy can also sustain high CMOS backend processing temperatures.

## ACKNOWLEDGEMENTS

This research is supported by the National Research Foundation (NRF), Prime Minister's Office, Singapore, under its Competitive Research Programme (CRP Award No. NRF CRP12-2013-01).

Figure Captions

FIG. 1. (a) 3D illustration of ST-FMR device with a schematic of measurement setup. (b) Optical microscope image of a ST-FMR device. The orange circle and dotted lines in (b) indicate the corresponding section of device illustrated in (a). (c) Representative ST-FMR spectra measured from Py (5 nm)/$Cu_{1-x}Pt_x$ (6 nm) bilayer for x = 0, 6.6, 13.7, and 19.7% for an applied microwave power of 16 dBm and a microwave frequency of 8 GHz.

FIG. 2. Lorentzian fittings of ST-FMR spectra from Py (5 nm)/$Cu_{1-x}Pt_x$ (6 nm) bilayer for x = 0, 6.6, 13.7, and 19.7% in the negative $H_{ext}$ range for an applied microwave power of 16 dBm and a microwave frequency of 8 GHz. As the Pt concentration in the CuPt alloy increases, the spin Hall torque from the CuPt layer increases which is indicated by an increase in the amplitude of the symmetric component (red curve).

FIG. 3. (a) $\theta_{SH}$ for different Pt concentrations extracted from ST-FMR spectra by using the $V_S / V_A$ ratio method (blue circles) and only $V_S$ method (red squares). The quantitative agreement of $\theta_{SH}$ from both the methods suggests that the field-like torque from spin-orbit effects is negligible. (b) Effective Gilbert damping ($\alpha_{eff}$) extracted from the ST-FMR linewidth as a function of different Pt concentrations. (c) $\rho_{CuPt}$ for different Pt concentrations (blue symbols) with a fit (red curve) using the Nordhiem rule. Most of the data points fit well suggesting that the CuPt alloy is homogeneous.

FIG. 4. (a) $\theta_{SH}$ as a function of $\rho_{CuPt}$ in the Cu-rich regime. (b) Extracted values of $\left| \rho_{SH}^{imp} \right|$ plotted against $\rho_{imp}$. Blue line indicates a fit to the data.

FIG. 5. (a) $\theta_{SH}$ as a function of CuPt thickness for x = 3.5, 13.7 and 19.7%. The data are fitted for each composition (dotted lines) to extract the values of $\theta_{SH}^0$ and $\lambda_{SH}$. (b) $\lambda_{SH}$ plotted (blue circles) against the conductivity of the CuPt , $\sigma$. The linear fitting (red line) of the data suggests that the spin relaxation in CuPt arises possibly from Elliot-Yafet mechanism.

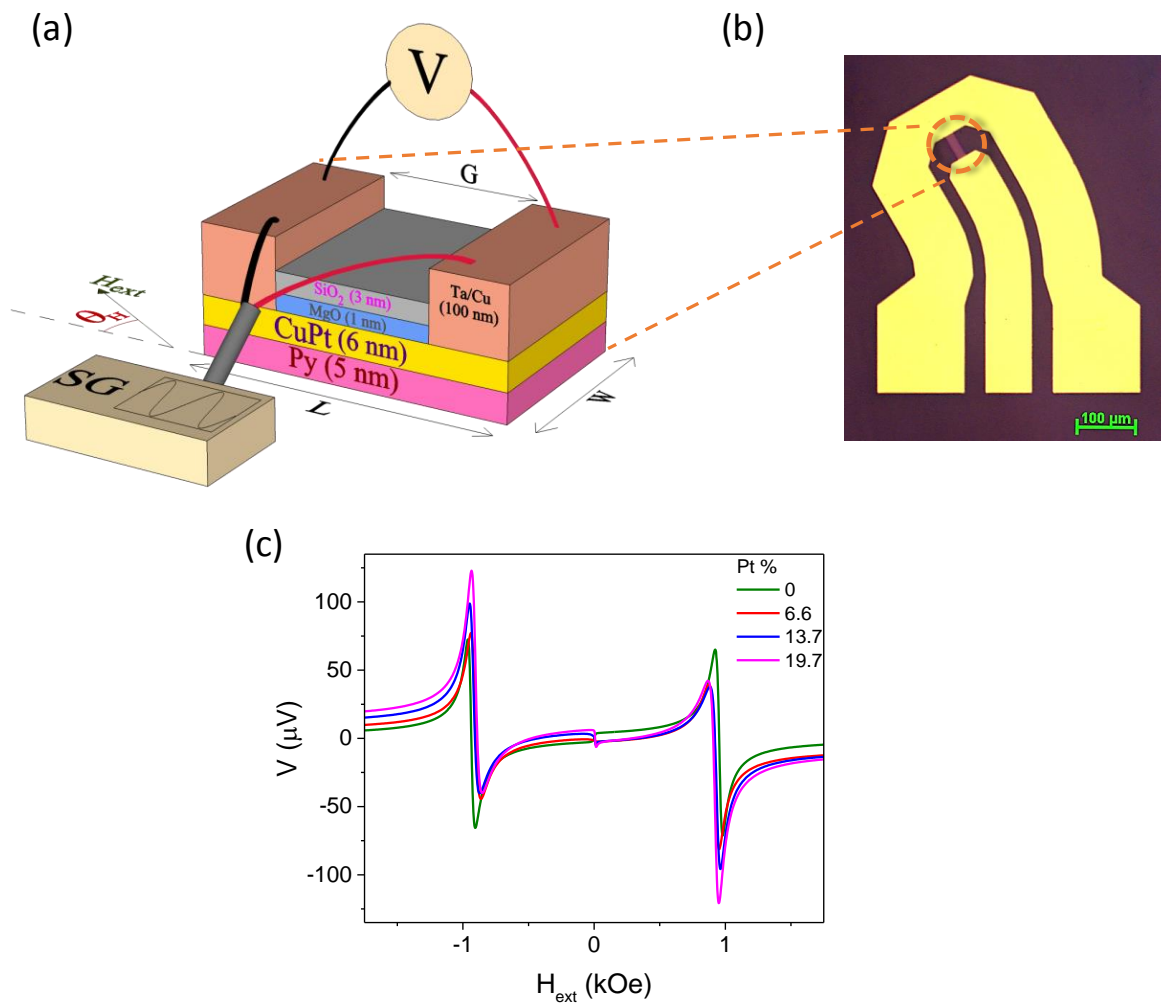

(a)

(b)

(c)

Figure 1

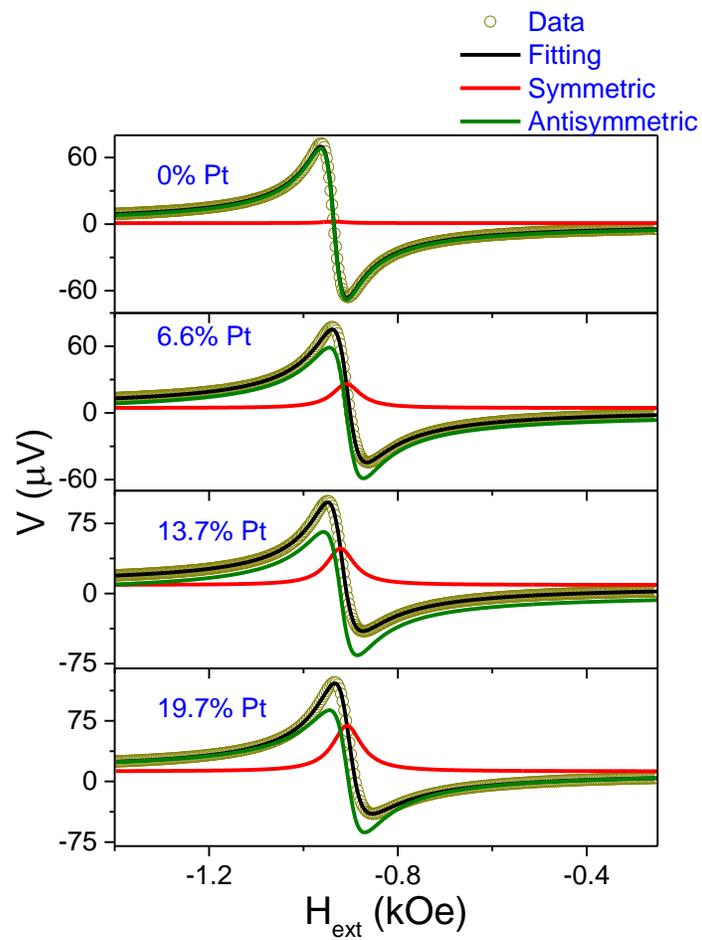

Figure 2

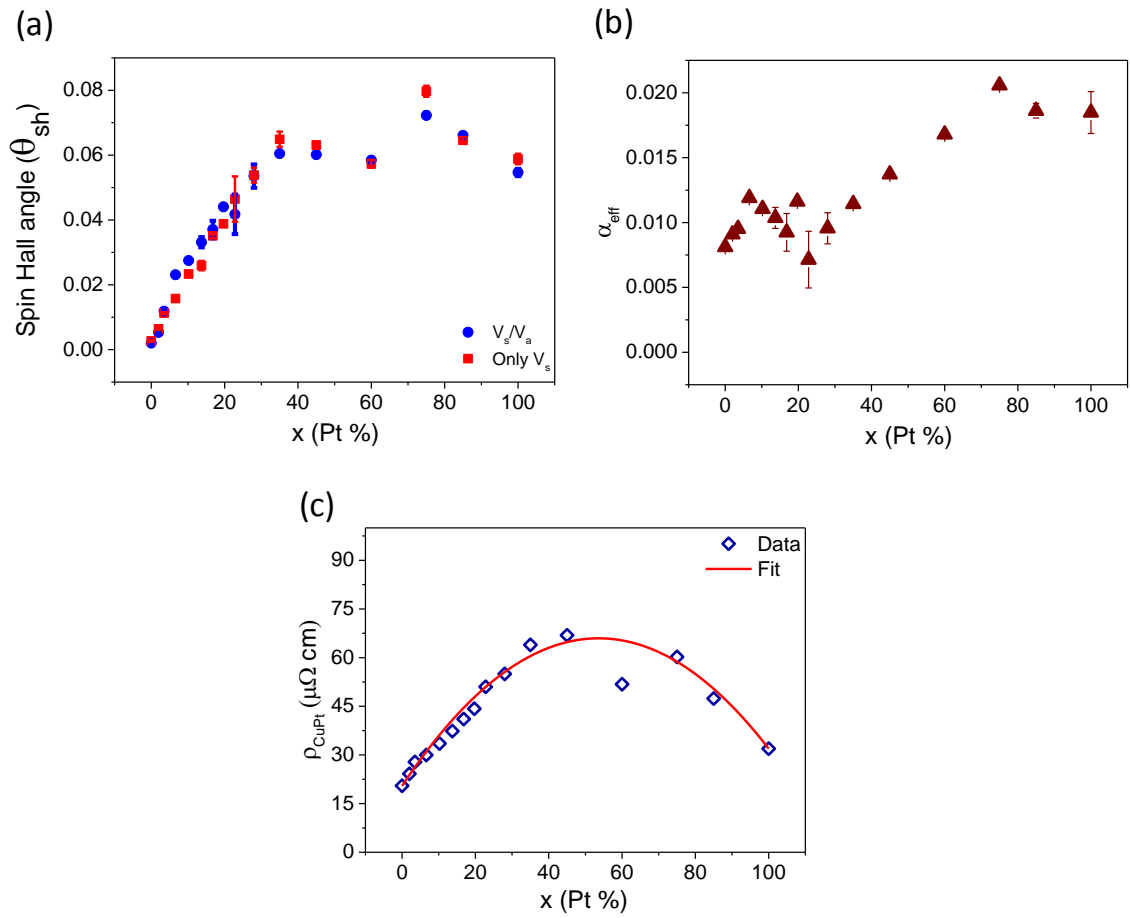

Figure 3

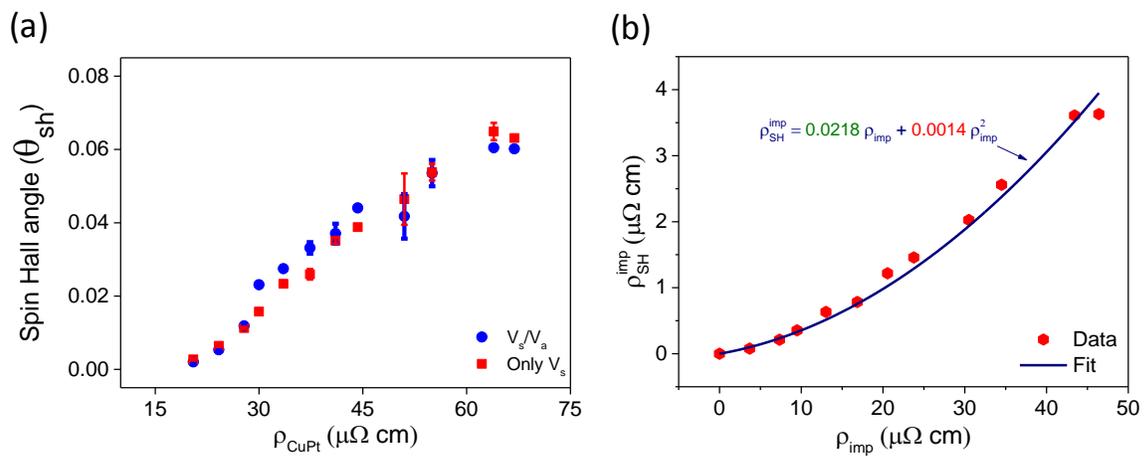

(a)

(b)

$\rho_{SH}^{imp} = 0.0218\,\rho_{imp} + 0.0014\,\rho_{imp}^2$

Figure 4

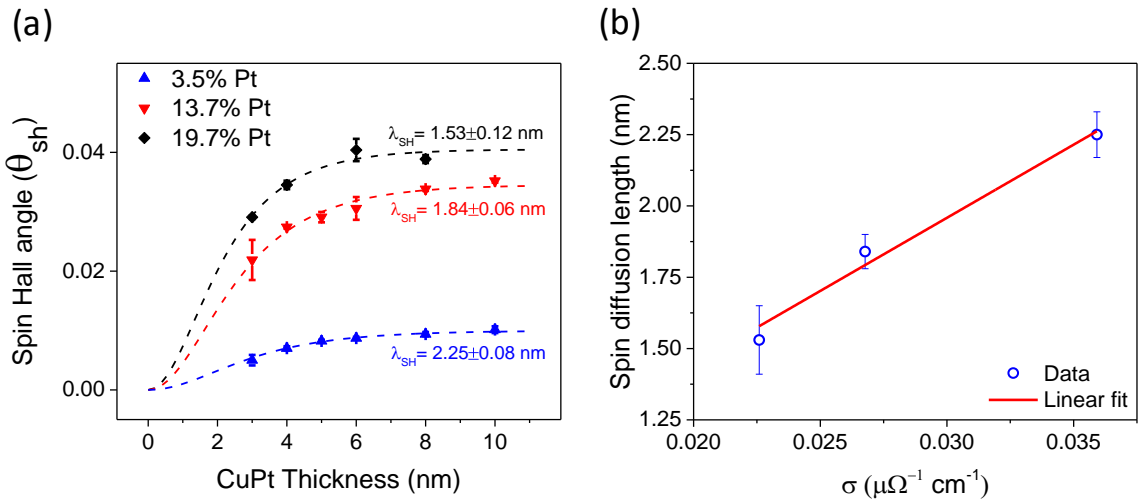

Figure 5

TABLE I. Estimated values of $\theta_{SH}^0$, $\lambda_{SH}$ and the product $\theta_{SH}^0 \lambda_{SH}$ for the CuPt alloy for x = 3.5, 13.7, and 19.7%. The parameters for CuBi, CuPb and CuIr alloys from literature are also shown for comparison.

| Alloy | $\theta_{SH}^0$ | $\lambda_{SH}$ (nm) | $\theta_{SH}^0 \lambda_{SH}$ (nm) | Ref. |
|---|---|---|---|---|
| $Cu_{96.5}Pt_{3.5}$ (300 K) | 0.010 | $2.25 \pm 0.08$ | 0.02 | This work |
| $Cu_{86.3}Pt_{13.7}$ (300 K) | 0.034 | $1.84 \pm 0.06$ | 0.06 | This work |
| $Cu_{80.3}Pt_{19.7}$ (300 K) | 0.040 | $1.53 \pm 0.12$ | $0.06 \pm 0.01$ | This work |
| $Cu_{99.5}Bi_{0.5}$ (10 K) | $-0.24 \pm 0.09$ | $45 \pm 14$ | $-11 \pm 5.3$ | [20,21] |
| $Cu_{99.5}Pb_{0.5}$ (10 K) | $-0.13 \pm 0.03$ | $53 \pm 15$ | $-7 \pm 2.5$ | [21] |
| $Cu_{99}Ir_1$ (10 K) | $0.021 \pm 0.006$ | $36 \pm 7$ | $0.8 \pm 0.27$ | [19,21] |